\documentclass[article]{emulateapj}

\newcommand{\hi}{\mbox{\rm \ion{H}{1}}}

\newcommand{\kmpers}{\mbox{km~s$^{-1}$}}

\newcommand{\xcounits}{\mbox{cm$^{-2}$ (K km s$^{-1}$)$^{-1}$}}

\newcommand{\xco}{\mbox{$X_{\rm CO}$}}

\begin{document}
\title{The Low CO Content of the Extremely Metal Poor Galaxy I~Zw~18}

\author{Adam Leroy\altaffilmark{1}, John Cannon\altaffilmark{1,2}, Fabian
  Walter\altaffilmark{1}, Alberto Bolatto\altaffilmark{3}, Axel
  Weiss\altaffilmark{4}}

\altaffiltext{1}{Max-Planck-Institut f\"{u}r Astronomie, K\"{o}nigstuhl 17,
  D-69117, Heidelberg, Germany; email: {\tt leroy@mpia-hd.mpg.de}}

\altaffiltext{2}{Astronomy Department, Wesleyan University, Middletown, CT
  06459, {\tt cannon@astro.wesleyan.edu}}

\altaffiltext{3}{Radio Astronomy Lab, UC Berkeley, 601 Campbell Hall,
 Berkeley, CA, 94720}

\altaffiltext{4}{MPIfR, Auf dem H\"{u}gel 69, 53121, Bonn, Germany}

\begin{abstract}
  We present sensitive molecular line observations of the metal-poor blue
  compact dwarf I~Zw~18 obtained with the IRAM Plateau de Bure interferometer.
  These data constrain the CO $J=1\rightarrow0$ luminosity within our 300~pc
  (FWHM) beam to be $L_{\rm{CO}} < 1 \times 10^5$~K~km~s$^{-1}$~pc$^2$
  ($I_{\rm{CO}} < 1$~K~km~s$^{-1}$), an order of magnitude lower than previous
  limits.  Although I~Zw~18 is starbursting, it has a CO luminosity similar to
  or less than nearby low-mass irregulars (e.g. NGC~1569, the SMC, and
  NGC~6822).  There is less CO in I~Zw~18 relative to its $B$-band luminosity,
  \hi\ mass, or star formation rate than in spiral or dwarf starburst galaxies
  (including the nearby dwarf starburst IC~10).  Comparing the star formation
  rate to our CO upper limit reveals that unless molecular gas forms stars
  much more efficiently in I~Zw~18 than in our own galaxy, it must have a very
  low CO-to-H$_2$ ratio, $\sim 10^{-2}$ times the Galactic value. We detect
  3mm continuum emission, presumably due to thermal dust and free-free
  emission, towards the radio peak.
\end{abstract}


\section{Introduction}
\label{INTRO}

With the lowest nebular metallicity in the nearby universe \citep[$12 + \log
O/H \approx 7.2$,][]{SKILLMAN93}, the blue compact dwarf I~Zw~18 plays an
important role in our understanding of galaxy evolution. Vigorous ongoing star
formation implies the presence of molecular gas, but direct evidence has been
elusive. \citet{VIDALMADJAR00} showed that there is not significant diffuse
H$_2$, but \citet{CANNON02} found $\sim 10^3$ M$_{\odot}$ of dust organized in
clumps with sizes 50 -- 100~pc. \citet{VIDALMADJAR00} did not rule out
compact, dense molecular clouds, and \citet{CANNON02} argued that this dust
may indicate the presence of molecular gas.

Observations by \citet{ARNAULT88} and \citet{GONDHALEKAR98} failed to detect
CO $J=1\rightarrow0$ emission, the most commonly used tracer of H$_2$. This is
not surprising. The low dust abundance and intense radiation fields found in
I~Zw~18 may have a dramatic impact on the formation of H$_2$ and structure of
molecular clouds. A large fraction of the H$_2$ may exist in extended
envelopes surrounding relatively compact cold cores. In these envelopes, H$_2$
self-shields while CO is dissociated \citep{MALONEY88}. The result may be that
in such galaxies [CII] or FIR emission trace H$_2$ better than CO
\citep[][]{MADDEN97,ISRAEL97A,PAK98}.  Further, H$_2$ may simply be
underabundant, as there is a lack of grains on which to form while
photodissociation is enhanced by an intense UV field.  Indeed, \citet{BELL06}
found that at $Z = Z_{\odot}/100$, a molecular cloud may take as long as a Gyr
to reach chemical equilibrium.

A low CO content in I~Zw~18 is then expected, and a stringent upper limit
would lend observational support to predictions for molecular cloud structure
at low metallicity. However, while the existing upper limits are sensitive in
an absolute sense, they do not even show I~Zw~18 to have a lower {\em
  normalized} CO content than a spiral galaxy (e.g. less CO per $B$-band
luminosity). The low luminosity \citep[$M_B \approx -14.7$,][]{GILDEPAZ03} and
large distance \citep[d=14~Mpc,][]{IZOTOV04} of this system require very
sensitive observations to set a meaningful upper limit.

In this letter we present observations, obtained with the IRAM Plateau de Bure
Interferometer (PdBI)\footnote{Based on observations carried out with the IRAM
  Plateau de Bure Interferometer. IRAM is supported by INSU/CNRS (France), MPG
  (Germany) and IGN (Spain)."}, that constrain the CO luminosity,
$L_{\rm{CO}}$, to be equal to or less than that of nearby CO-poor
(non-starbursting) dwarf irregulars.

\section{Observations}
\label{DATA}

I~Zw~18 was observed with the IRAM Plateau de Bure Interferometer on 17, 21,
and 27 April and 13 May 2004 for a total of 11 hours. The phase calibrators
were 0836+710 ($F_{\nu} (\rm{115 GHz}) \approx 1.1$~Jy), and 0954+556
($F_{\nu} (\rm{115 GHz}) \approx 0.35$~Jy). One or more calibrators with known
fluxes were also observed during each track. The data were reduced at the IRAM
facility in Grenoble using the GILDAS software package; maps were prepared
using AIPS. The final CO $J=1\rightarrow0$ data cube has beam size $5.59''
\times 3.42''$, and a velocity (frequency) resolution of $6.5$~\kmpers\
($2.5$~MHz).  The velocity coverage stretches from $v_{LSR} \approx 50$ to
$1450$~\kmpers. The data have an RMS noise of $3.77$~mJy~beam$^{-1}$ ($18$~mK;
$1$ Jy beam$^{-1}$ = $4.8$ K).  The $44''$ (FWHM) primary beam completely
covers the galaxy.  Based on variation of the relative fluxes of the
calibrators, we estimate the gain uncertainty to be $< 15\%$.

\section{Results}
\label{RESULTS}

\subsection{Upper Limit on CO Emission}

To search for significant CO emission, we smooth the cube to 20~\kmpers\
velocity resolution, a typical line width for CO at our spatial resolution
\citep[e.g.,][]{HELFER03}. The noise per channel map in this smoothed cube is
$\sigma_{20} \approx 0.25$~K~km~s$^{-1}$.  Over the \hi\ velocity range
\citep[710 -- 810 \kmpers,][]{VANZEE98}, there are no regions with
$I_{\rm{CO},20} > 1$~K~km~s$^{-1}$ (4$\sigma$) within the primary beam. We
pick a slightly conservative upper limit for two reasons.  First, if there
were CO emission with this intensity we would be certain of detecting it.
Second, the noise in the cube is slightly non-Gaussian, so that the false
positive rate for $I_{\rm{CO},20} > 1$~K~km~s$^{-1}$ --- estimated from the
negatives and the channel maps outside the \hi\ velocity range --- is $\sim
0.2$\%, very close to that of a $3\sigma$ deviate.

For $d = 14$~Mpc, the synthesized beam has a FWHM of 300~pc and an area of
$1.0 \times 10^5$~pc$^2$. Our intensity limit, $I_{\rm{CO}} <
1$~K~km~s$^{-1}$, therefore translates to a CO luminosity limit of
$L_{\rm{CO}} < 1 \times 10^5$~K~km~s$^{-1}$~pc$^2$.

\begin{figure}
\begin{center}
  \plotone{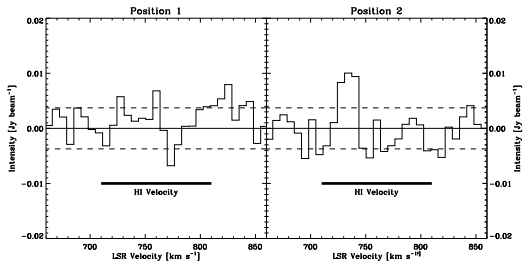} \figcaption{\label{SPECS} CO $1\rightarrow0$ spectra
    of I~Zw~18 towards the radio continuum/H$\alpha$ peak (left) and the
    highest significance spectra (right), which
    is still too faint to classify as more than marginal. The locations of
    both spectra are shown in Figure \ref{CONTMAP}. Dashed horizontal lines
    show the magnitude of the RMS noise.}
\end{center}
\end{figure}

There is a marginal signal toward the southern knot of H$\alpha$ emission
($9^{\rm h}34^{\rm m}02^{\rm s}.4$, $55^{\circ}14'23''.0$). This emission has
the largest $|I_{\rm{CO},20}|$ found over the \hi\ velocity range,
corresponding to $L_{\rm{CO}} \sim 8 \times 10^4$~K~km~s$^{-1}$~pc$^2$, just
below our limit. This same line of sight also shows $|I_{CO}| > 2\sigma$ over
three consecutive channels, a feature seen along only one other line of sight
(in negative) over the \hi\ velocity range.  The marginal signal is
suggestively located in the southeast of I~Zw~18, where \citet{CANNON02}
identified several potential sites of molecular gas from regions of relatively
high extinction.  While tantalizing, the signal is not strong enough to be
categorized as a detection. Figure \ref{SPECS} shows CO spectra towards the
H$\alpha$/radio continuum peak \citep[][see Figure
\ref{CONTMAP}]{CANNON02,CANNON05,HUNT05A} and this marginal signal.

\subsection{Continuum Emission}

\begin{figure*}
\begin{center}
  \plotone{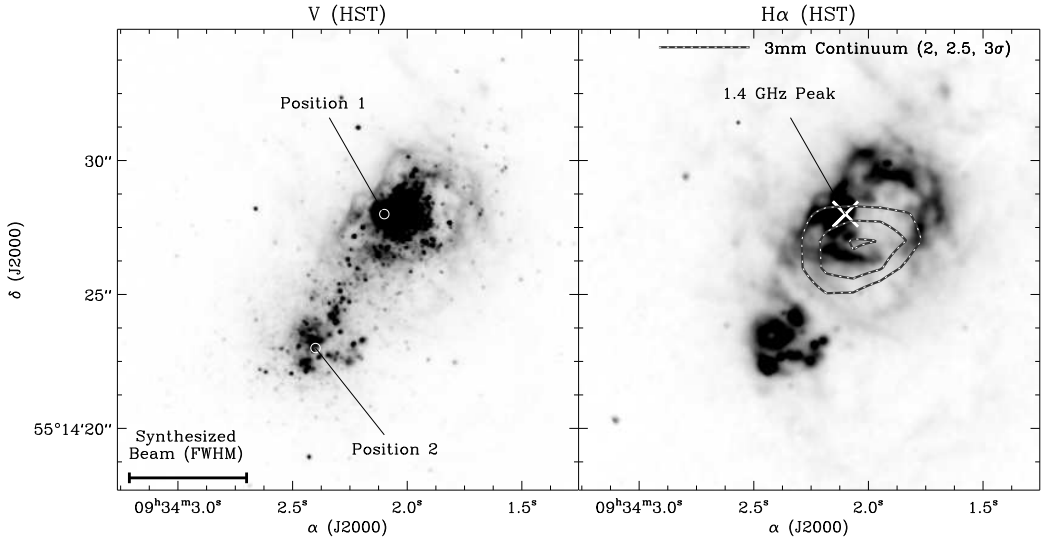} \figcaption{\label{CONTMAP} $V$-band and
    H$\alpha$ \citep[right,][]{CANNON02} images of I~Zw~18. Overlays on the
    left image show the size of the synthesized beam and the locations of the
    spectra shown in Figure \ref{SPECS}.  Contours on the right image show
    continuum emission in increments of $0.5\sigma$ significance and the
    location of the radio continuum peak. The primary beam is larger than the
    area shown.  Both optical maps are on linear stretches. $V$-band data
    obtained from the MAST Archive, originally observed for GO program 9400,
    PI: T. Thuan).}
\end{center}
\end{figure*}

We average the data over all channels and produce a continuum map with noise
$\sigma_{115\rm{GHz}} = 0.35$ mJy beam$^{-1}$. The highest value in the map is
$I_{115\rm{GHz}} = 1.06 \pm 0.35$ mJy beam$^{-1}$ at $\alpha_{2000} = 9^{\rm
  h}34^{\rm m}02^{\rm s}.1$, $\delta_{2000}=+55^{\circ}~14'~27''.0$. This is
within a fraction of a beam of the 1.4~GHz peak identified by \citet[][
$\alpha_{2000} = 9^{\rm h}34^{\rm m}02^{\rm s}.1$,
$\delta_{2000}=+55^{\circ}~14'~28''.06$]{CANNON05} and \citet[][$\alpha_{2000}
= 9^{\rm h}34^{\rm m}02^{\rm s}$,
$\delta_{2000}=+55^{\circ}~14'~29''.06$]{HUNT05A}. Figure \ref{CONTMAP} shows
the radio continuum peak and 115 GHz continuum contours plotted over H$\alpha$
emission from I~Zw~18 \citep{CANNON02}.  There is only one other region with
$|I_{115\rm{GHz}}| > 3\sigma_{115\rm{GHz}}$ within the primary beam and the
star-forming extent of I~Zw~18 occupies $\approx 10~\%$ of the primary beam.
Therefore, we estimate the chance of a false positive coincident with the
galaxy to be only $\sim 10\%$.

\section{Discussion}
\label{DISCUSSION}

Here we discuss the implications of our CO upper limit and continuum
detection. We adopt the following properties for I~Zw~18, all scaled to
$d=14$~Mpc: $M_B = -14.7$ \citep{GILDEPAZ03}, $M_{HI} = 1.4 \times
10^8$~M$_{\odot}$ \citep{VANZEE98}, H$\alpha$ luminosity $\log_{10} H_{\alpha}
= 39.9$ erg s$^{-1}$ \citep{CANNON02,GILDEPAZ03}, 1.4~GHz flux $F_{1.4} =
1.79$~mJy \citep{CANNON05}.

\subsection{Point Source Luminosity}

Our upper limit along each line of sight, $L_{\rm{CO}} < 1 \times
10^5$~K~km~s$^{-1}$~pc$^2$, matches the luminosity of a fairly massive
Galactic giant molecular cloud \citep[][]{BLITZ93}. For a Galactic CO-to-H$_2$
conversion factor, $2 \times 10^{20}$~\xcounits, the corresponding molecular
gas mass is $M_{Mol} \approx 4.4 \times 10^5$~M$_{\odot}$, similar to the mass
of the Orion-Monoceros complex \citep[e.g.][]{WILSON05}.

\subsection{Comparison With More Luminous Galaxies}
\begin{figure}
\begin{center}
  \plotone{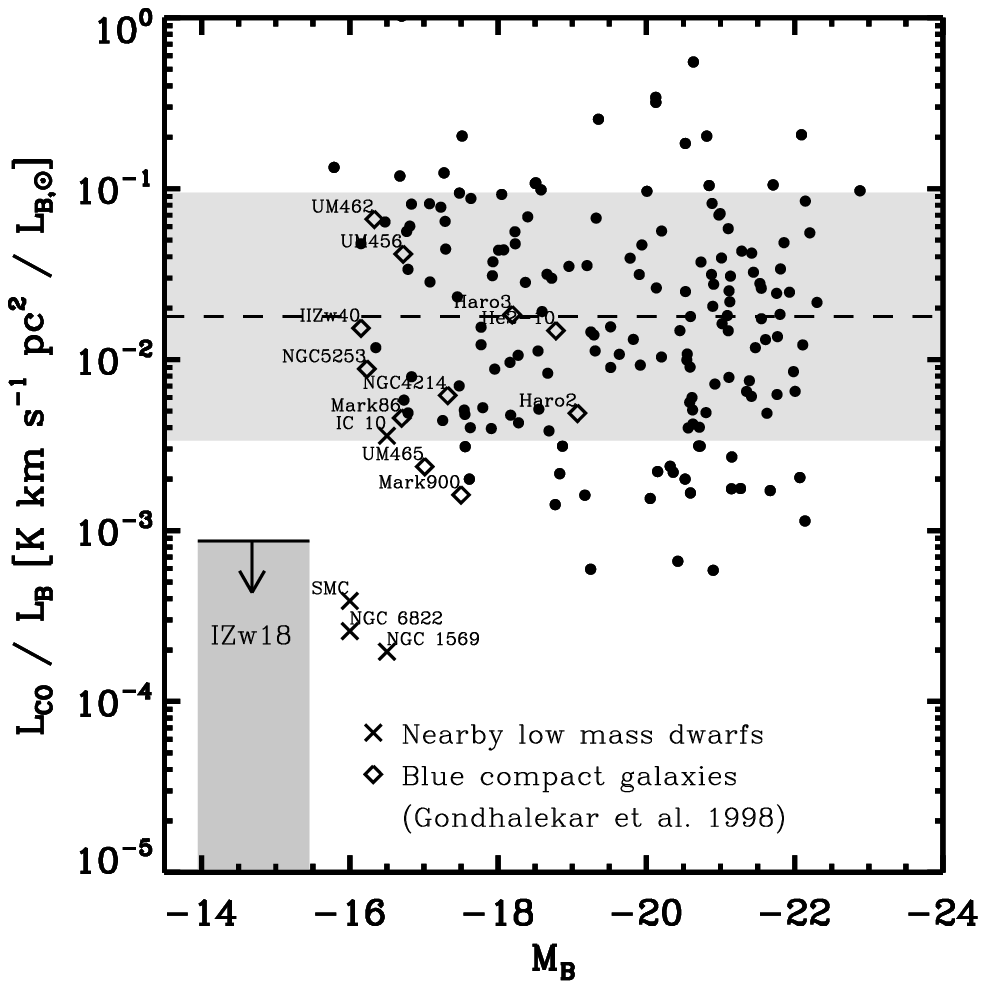} \figcaption{\label{NORMPLOT} CO luminosity normalized by
    absolute blue magnitude for galaxies with Hubble Type Sb or later
    \citep[black circles,][]{YOUNG95,ELFHAG96,BOKER03,LEROY05}. We also plot
    nearby dwarfs from Table \ref{NEARDWARFTAB} (crosses) and blue compact
    galaxies compiled by \citet[][, diamonds]{GONDHALEKAR98}. The shaded
    regions shows our upper limit for I~Zw~18, with the range in $M_B$ for
    distances from 10 to 20~Mpc.  The dashed line and light shaded region show
    the median value and $1\sigma$ scatter in $L_{\rm{CO}}/L_B$ for spirals
    and dwarf starbursts.  {\em Methodology:} We extrapolate from
    $I_{\rm{CO}}$ in central pointings to $L_{\rm{CO}}$ assuming the CO to
    have an exponential profile with scale length 0.1 $d_{25}$
    \citep{YOUNG95}, including only galaxies where the central pointing
    measures $> 20\%$ of $L_{CO}$. We adopt $B$ magnitudes (corrected for
    internal and Galactic extinction), distances (Tully-Fisher when available,
    otherwise Virgocentric-flow corrected Hubble flow), and radii from LEDA
    \citep{PATUREL03}.}
\end{center}
\end{figure}

In galaxies detected by CO surveys, the CO content per unit $B$-band
luminosity is fairly constant. Figure \ref{NORMPLOT} shows the CO luminosity
normalized by $B$-band luminosity, $L_{\rm{CO}}/L_B$, as a function of
absolute $B$-band magnitude ($L_{B}$ is extinction corrected).
$L_{\rm{CO}}/L_B$ is nearly constant over two orders of magnitude in $L_{B}$,
though with substantial scatter (much of it due to the extrapolation from a
single pointing to $L_{\rm{CO}}$).

Based on these data and assuming that $L_{\rm{CO}}$ is not a function of the
metallicity of the galaxy, we may extrapolate to an expected CO luminosity for
I~Zw~18. For $M_{B,\rm{IZw18}} \approx -14.7$ the CO luminosity corresponding
to the median value of $L_{\rm{CO}}/L_B$ (dashed line) in Figure
\ref{NORMPLOT} is $L_{\rm{CO,IZw18}} \approx 1.7 \times
10^6$~K~km~s$^{-1}$~pc$^2$.  The H$\alpha$, 1.4~GHz, and \hi\ luminosities
lead to similar predictions. \citet{YOUNG96} found $M_{H2}/L_{H\alpha} \approx
10 L_{\odot}/M_{\odot}$ for Sd--Irr galaxies, which implies $L_{\rm{CO,IZw18}}
\sim 4 \times 10^6$~K~km~s$^{-1}$~pc$^2$. \citet{MURGIA05} measured
$F_{\rm{CO}}/F_{1.4} \approx 10$~Jy~km~s$^{-1}$ (mJy)$^{-1}$ for spirals, that
would imply L$_{\rm{CO,IZw18}} \sim 10^7$~K~km~s$^{-1}$. For Sd/Sm galaxies,
$M_{H2}/M_{HI} \approx 0.2$ \citep{YOUNG91}, leading to $L_{\rm{CO,IZw18}}
\sim 5 \times 10^6$~K~km~s$^{-1}$~pc$^2$. Both $M_{H2}/L_{H\alpha}$ and
$M_{H2}/M_{HI}$ tend to be even higher in earlier-type spirals.

Therefore, surveys would predict $L_{\rm{CO,IZw18}} \gtrsim 2 \times
10^6$~K~km~s$^{-1}$~pc$^2$, very close to the previously established upper
limits of $2 - 3 \times 10^6$~K~km~s$^{-1}$pc$^2$
\citep{ARNAULT88,GONDHALEKAR98}. With the present observations, we constrain
$L_{\rm{CO}} < 1 \times 10^5$~K~km~s$^{-1}$pc$^2$ and thus clearly rule out
$L_{\rm{CO}} \sim 10^6$~K~km~s$^{-1}$~pc$^2$. This may be seen in Figure
\ref{NORMPLOT}; even if I~Zw~18 has the highest possible CO content, it will
still have a lower $L_{\rm{CO}}/L_B$ than $97\%$ of the survey galaxies.

\subsection{Comparison With Nearby Metal-Poor Dwarfs}

The subset of irregular galaxies detected by CO surveys tend to be CO-rich and
actively star-forming, resembling scaled-down versions of spiral galaxies
\citep{YOUNG95,YOUNG96,LEROY05}. Such galaxies may not be representative of
all dwarfs. Because they are nearby, several of the closest dwarf irregulars
have been detected despite very small $L_{\rm{CO}}$. With their low masses and
metallicities, they may represent good points of comparison for I~Zw~18.
Table~\ref{NEARDWARFTAB} and Figure \ref{NORMPLOT} show CO luminosities and
$L_{\rm{CO}}/L_B$ for four nearby dwarfs: NGC~1569, the Small Magellanic Cloud
(SMC), NGC~6822, and IC~10. The SMC, NGC~1569, and NGC~6822 have $L_{\rm{CO}}
\sim 10^5$~K~km~s$^{-1}$~pc$^2$, close to our upper limit, and occupy a region
of $L_{\rm{CO}}/L_B$-$L_B$ parameter space similar to I~Zw~18. All four of
these galaxies have active star formation but very low CO content relative to
their other properties.

We test whether our observations would have detected CO in NGC~1569, the SMC,
and IC~10 at the plausible lower limit of 10~Mpc (from $H_0 = 72$ km s$^{-1}$)
or our adopted distance of 14~Mpc. We convolve the integrated intensity maps
to resolutions of 210 and 300~pc and measure the peak integrated intensity.
The results appear in columns 4 and 5 of Table~\ref{NEARDWARFTAB}. The PdBI
observations of NGC~1569
resolve out most of the flux, so we also apply this test to a distribution
with the size and luminosity derived by \citet{GREVE96} from single dish
observations. Our observations would detect an analog to IC~10 but not the
SMC, with NGC~1569 an intermediate case.  With a factor of $\sim 3$ better
sensitivity (requiring $\sim 10$ times more observing time) we would expect to
detect all three nearby galaxies.  However, achieving such sensitivity with
present instrumentation will be quite challenging. ALMA will likely be
necessary to place stronger constraints on CO in galaxies like I~Zw~18.

IC~10 may be the nearest blue compact dwarf \citep{RICHER01}, so it may be
telling that we would detect it at the distance of I~Zw~18. The blue compact
galaxies that have been detected in CO have L$_{\rm{CO}}$/L$_B$ similar to
IC~10 \citep[][the diamonds in Figure \ref{NORMPLOT}]{GONDHALEKAR98}. Most
searches for CO towards BCDs have yielded nondetections, so those detected may
not be representative, but I~Zw~18 is clearly not among the ``CO-rich''
portion of the BCD population.

\subsection{Interpretation of the Continuum}

We measure continuum intensity of $F_{115\rm{GHz}} = 1.06 \pm 0.35$ mJy
towards the radio continuum peak. The continuum is detected along only one
line of sight, so we refer to it here as a point source and compare it to
integrated values for I Zw 18.  $F_{115\rm{GHz}}$ is expected to be the
product of mainly two types of emission: thermal free-free emission and
thermal dust emission. At long wavelengths, the integrated thermal free-free
emission is $F_{1.4\rm{GHz}} (\rm{free-free}) \approx 0.52$ -- $0.75$~mJy
\citep{CANNON05,HUNT05A}, implying $F_{115\rm{GHz}} (\rm{free-free}) = 0.36$
-- $0.51$~mJy at $115$~GHz ($F_{\nu} \propto \nu^{-0.1}$). The H$\alpha$ flux
predicts a similar value, $F_{115\rm{GHz}} (\rm{free-free}) = 0.34$~mJy
\citep[][Equation 1]{CANNON05}.  \citet{HUNT05B} placed an upper limit of
$F_{\nu} (850) < 2.5$~mJy on dust continuum emission at 850$\mu$m; this is
consistent with the $\sim 5 \times 10^3$~M$_{\odot}$ estimated by
\citet[][]{CANNON02} given almost any reasonable dust properties.
Extrapolating this to $2.6$~mm assuming a pure blackbody spectrum, the
shallowest plausible SED, constrains thermal emission from dust to be $<
0.25$~mJy at $115$~GHz. Based on these data, we would predict $F_{115\rm{GHz}}
\lesssim 0.75$~mJy. Thus our measured $F_{115\rm{GHz}}$ is consistent with,
but somewhat higher than, the thermal free-free plus dust emission expected
based on optical, centimeter, and submillimeter data.

\subsection{Relation to Star Formation}

I~Zw~18 has a star formation rate $\sim 0.06$ -- $0.1$ M$_{\odot}$ yr$^{-1}$,
based on H$\alpha$ and cm radio continuum measurements
\citep{CANNON02,KENNICUTT98A,HUNT05A}.  Our continuum flux suggests a slightly
higher value $\approx 0.15$ -- $0.2$~M$_{\odot}$~yr$^{-1}$
\citep[following][]{HUNT05A,CONDON92}, with the exact value depending on the
contribution from thermal dust emission. For any value in this range, the star
formation rate per CO luminosity, $SFR/L_{\rm{CO}}$ is much higher in I~Zw~18
than in spirals. For comparison, our upper limit and the molecular ``Schmidt
Law'' derived by \citet{MURGIA02} predicts a star formation rate $\lesssim 2
\times 10^{-4}$ M$_{\odot}$ yr$^{-1}$. Fits by \citet{YOUNG96} and
\citet[][applied to just the molecular limit]{KENNICUTT98B} yield similar
values. Again, I~Zw~18 is similar to the SMC and NGC~6822, which have star
formation rates of $0.05$ M$_{\odot}$ yr$^{-1}$ and $0.04$ M$_{\odot}$
yr$^{-1}$ \citep[][]{WILKE04,ISRAEL97B} and $L_{\rm{CO}} \sim
10^5$~K~km~s$^{-1}$~pc$^2$.

\begin{deluxetable*}{l c c c c l}
\tabletypesize{\small} 
\tablewidth{0pt} 
\tablecolumns{5}
\tablecaption{\label{NEARDWARFTAB} CO in Nearby Low Mass Galaxies}

\tablehead{\colhead{Galaxy} & \colhead{$M_{B}$} &
\colhead{$L_{\rm{CO}}$} & \colhead{$I_{CO,210}$\tablenotemark{a}} &
\colhead{$I_{CO,300}$\tablenotemark{a}} &
\colhead{Reference} \\
& (mag) & (K~km~s$^{-1}$~pc$^2$) & (K~km~s$^{-1}$) & (K~km~s$^{-1}$) & }

\startdata
NGC~1569 & $-16.5$ & $1.2 \times 10^5$ & $1.1$ & $0.8$ & \citet{GREVE96} \\
 & $-16.5$ & $0.2 \times 10^5$ & $0.8$ & $0.5$ & \citet{TAYLOR99} \\
SMC & $-16$ & $1.5 \times 10^5$ & $0.5$ & $0.4$ & \citet{MIZUNO01,MIZUNO06} \\
NGC~6822 & $-16$ & $1.2 \times 10^5$ & \nodata & \nodata & \citet{ISRAEL97B} \\
IC~10 & $-16.5$ & $2.2 \times 10^6$ & $3.8$ & $2.2$ & \citet{LEROY06} \\
\hline
I~Zw~18 & $-14.7$ & $ < 2 \times 10^6$ & \nodata & \nodata & \citet{ARNAULT88,GONDHALEKAR98} \\
I~Zw~18 & $-14.7$ & $ \lesssim 1 \times 10^5$ & $< 1$ & $< 1$ & this paper
\enddata
\tablenotetext{a}{Peak integrated intensity at 210 and 300 pc,
  corresponding to our beam size at 10 and 14 Mpc, respectively.}
\end{deluxetable*}

\subsection{Variations in \xco}

Several calibrations of the CO-to-H$_2$ conversion factor, \xco\, as a
function of metallicity exist in the literature. The topic has been
controversial and these calibrations range from little or no dependence
\citep[e.g.][]{WALTER03,ROSOLOWSKY03} to very steep dependence \citep[e.g.,
$\xco\ \propto Z^{-2.7}$][]{ISRAEL97A}. Comparing the star formation rate to
our CO upper limit, we may rule out that I~Zw~18 has a Galactic \xco\ unless
molecular gas in I~Zw~18 forms stars much more efficiently than in the Galaxy.
Either the ratio of CO-to-H$_2$ is low in I~Zw~18 or molecular gas in this
galaxy forms stars with an efficiency two orders of magnitude higher than that
in spiral galaxies.

\section{Conclusions}
\label{CONCLUSIONS}

We present new, sensitive observations of the metal-poor dwarf galaxy I~Zw~18
at 3~mm using the Plateau de Bure Interferometer. These data constrain the
integrated CO $J=1\rightarrow0$ intensity to be $I_{\rm{CO}} <
1$~K~km~s$^{-1}$ over our $300$~pc (FWHM) beam and the luminosity to be
$L_{\rm{CO}} < 1 \times 10^5$~K~km~s$^{-1}$~pc$^2$.

I~Zw~18 has less CO relative to its $B$-band luminosity, \hi\ mass, or SFR
than spiral galaxies or dwarf starbursts, including more metal-rich blue
compact galaxies such as IC~10 \citep[$Z_{\rm{IC~10}} \sim
Z_{\odot}/4$,][]{LEE03}. Because of its small size and large distance, these
are the first observations to impose this constraint. 

We show that I~Zw~18 should be grouped with several local analogs ---
NGC~1569, the SMC, NGC~6822 --- as a galaxy with active star formation but a
very low CO content relative to its other properties. In these galaxies,
observations suggest that the environment affects the molecular gas and these
data suggest that the same is true in I~Zw~18. A simple comparison of star
formation rate to CO content shows that this must be true at a basic level:
either the ratio of CO to H$_2$ is dramatically low in I~Zw~18 or molecular
gas in this galaxy forms stars with an efficiency two orders of magnitude
higher than that in spiral galaxies.

We detect 3mm continuum with $F_{115~\rm{GHz}} = 1.06 \pm 0.35$ mJy coincident
with the radio peak identified by \citet{CANNON05} and \citet{HUNT05A}. This
flux is consistent with but somewhat higher than the thermal free-free plus
dust emission one would predict based on centimeter, submillimeter, and
optical measurements.

Finally, we note that improving on this limit with current instrumentation
will be quite challenging. The order of magnitude increase in sensitivity from
ALMA will be needed to place stronger constraints on CO in galaxies like
I~Zw~18.

\acknowledgements We thank Roberto Neri for his help reducing the data. We
acknowledge the usage of the HyperLeda database (http://leda.univ-lyon1.fr).


\begin{thebibliography}{}

\bibitem[Arnault et al.(1988)]{ARNAULT88} Arnault, P., Kunth, D., Casoli, F.,
  \& Combes, F.\ 1988, \aap, 205, 41

\bibitem[Bell et al.(2006)]{BELL06} Bell, T.~A., Roueff, E., Viti, S., \&
  Williams, D.~A.\ 2006, \mnras, 371, 1865

\bibitem[Blitz(1993)]{BLITZ93} Blitz, L.\ 1993, Protostars and Planets III,
  125

\bibitem[B{\"o}ker et al.(2003)]{BOKER03} B{\"o}ker, T., Lisenfeld, U., \&
  Schinnerer, E.\ 2003, \aap, 406, 87

\bibitem[Cannon et al.(2002)]{CANNON02} Cannon, J.~M., Skillman, 
E.~D., Garnett, D.~R., \& Dufour, R.~J.\ 2002, \apj, 565, 931

\bibitem[Cannon et al.(2005)]{CANNON05} Cannon, J.~M., Walter, F., Skillman,
  E.~D., \& van Zee, L.\ 2005, \apjl, 621, L21

\bibitem[Condon(1992)]{CONDON92} Condon, J.~J.\ 1992, \araa, 30, 575

\bibitem[Gil de Paz et al.(2003)]{GILDEPAZ03} Gil de Paz, A., Madore, B.~F.,
  \& Pevunova, O.\ 2003, \apjs, 147, 29

\bibitem[Elfhag et al.(1996)]{ELFHAG96} Elfhag, T., Booth, R.~S., Hoeglund,
  B., Johansson, L.~E.~B., \& Sandqvist, A.\ 1996, \aaps, 115, 439

\bibitem[Gondhalekar et al.(1998)]{GONDHALEKAR98} Gondhalekar, P.~M.,
  Johansson, L.~E.~B., Brosch, N., Glass, I.~S., \& Brinks, E.\ 1998, \aap,
  335, 152

\bibitem[Greve et al.(1996)]{GREVE96} Greve, A., Becker, R., Johansson,
  L.~E.~B., \& McKeith, C.~D.\ 1996, \aap, 312, 391

\bibitem[Helfer et al.(2003)]{HELFER03} Helfer, T.~T., Thornley, M.~D., Regan,
  M.~W., Wong, T., Sheth, K., Vogel, S.~N., Blitz, L., \& Bock, D.~C.-J.\
  2003, \apjs, 145, 259

\bibitem[Hunt et al.(2005a)]{HUNT05A} Hunt, L.~K., Dyer, K.~K., \& Thuan,
  T.~X.\ 2005a, \aap, 436, 837

\bibitem[Hunt et al.(2005b)]{HUNT05B} Hunt, L., Bianchi, S., \& Maiolino, R.\
  2005b, \aap, 434, 849

\bibitem[Israel(1997a)]{ISRAEL97A} Israel, F.~P.\ 1997, \aap, 328, 
471 

\bibitem[Israel(1997b)]{ISRAEL97B} Israel, F.~P.\ 1997, \aap, 317, 
65 

\bibitem[Izotov \& Thuan(2004)]{IZOTOV04} Izotov, Y.~I., \& Thuan, T.~X.\
  2004, \apj, 616, 768

\bibitem[Kennicutt(1998a)]{KENNICUTT98A} Kennicutt, R.~C., Jr.\ 1998a, \araa,
  36, 189

\bibitem[Kennicutt(1998b)]{KENNICUTT98B} Kennicutt, R.~C., Jr.\ 1998b, \apj,
  498, 541

\bibitem[Lee et al.(2003)]{LEE03} Lee, H., McCall, M.~L., \& Richer, M.~G.\
  2003, \aj, 125, 2975

\bibitem[Leroy et al.(2005)]{LEROY05} Leroy, A., Bolatto, A.~D., Simon, J.~D.,
  \& Blitz, L.\ 2005, \apj, 625, 763

\bibitem[Leroy et al.(2006)]{LEROY06} Leroy, A., Bolatto, A., Walter, F., \&
  Blitz, L.\ 2006, \apj, 643, 825

\bibitem[Madden et al.(1997)]{MADDEN97} Madden, S.~C., Poglitsch, A., Geis,
  N., Stacey, G.~J., \& Townes, C.~H.\ 1997, \apj, 483, 200

\bibitem[Maloney \& Black(1988)]{MALONEY88} Maloney, P., \& Black, J.~H.\
  1988, \apj, 325, 389

\bibitem[Mizuno et al.(2001)]{MIZUNO01} Mizuno, N., Rubio, M., Mizuno,
A., Yamaguchi, R., Onishi, T., \& Fukui, Y.\ 2001, \pasj, 53, L45

\bibitem[Mizuno et al.(2006)]{MIZUNO06} Mizuno, N., et al.\ 2006, in
prep.

\bibitem[Murgia et al.(2002)]{MURGIA02} Murgia, M., Crapsi, A., Moscadelli,
  L., \& Gregorini, L.\ 2002, \aap, 385, 412

\bibitem[Murgia et al.(2005)]{MURGIA05} Murgia, M., Helfer, T.~T., Ekers, R.,
  Blitz, L., Moscadelli, L., Wong, T., \& Paladino, R.\ 2005, \aap, 437, 389

\bibitem[Pak et al.(1998)]{PAK98} Pak, S., Jaffe, D.~T., van Dishoeck, E.~F.,
  Johansson, L.~E.~B., \& Booth, R.~S.\ 1998, \apj, 498, 735

\bibitem[Paturel et al.(2003)]{PATUREL03} Paturel, G., Petit, C., Prugniel,
  P., Theureau, G., Rousseau, J., Brouty, M., Dubois, P., \& Cambr{\'e}sy, L.\
  2003, \aap, 412, 45

\bibitem[Richer et al.(2001)]{RICHER01} Richer, M.~G., et al.\ 2001, \aap,
  370, 34

\bibitem[Rosolowsky et al.(2003)]{ROSOLOWSKY03} Rosolowsky, E., Engargiola,
  G., Plambeck, R., \& Blitz, L. \ 2003, \apj, 599, 258

\bibitem[Skillman \& Kennicutt(1993)]{SKILLMAN93} Skillman, E.~D., 
\& Kennicutt, R.~C., Jr.\ 1993, \apj, 411, 655 


\bibitem[Taylor et al.(1998)]{TAYLOR98} Taylor, C.~L., Kobulnicky, H.~A., \&
  Skillman, E.~D.\ 1998, \aj, 116, 2746

\bibitem[Taylor et al.(1999)]{TAYLOR99} Taylor, C.~L., H{\"u}ttemeister, S.,
  Klein, U., \& Greve, A.\ 1999, \aap, 349, 424

\bibitem[van Zee et al.(1998)]{VANZEE98} van Zee, L., Westpfahl, D., Haynes,
  M.~P., \& Salzer, J.~J.\ 1998, \aj, 115, 1000

\bibitem[Vidal-Madjar et al.(2000)]{VIDALMADJAR00} Vidal-Madjar, A., et al.\
  2000, \apjl, 538, L77


\bibitem[Walter(2003)]{WALTER03} Walter, F.\ 2003, IAU Symposium, 221, 176P

\bibitem[Wilke et al.(2004)]{WILKE04} Wilke, K., Klaas, U., Lemke, D.,
  Mattila, K., Stickel, M., \& Haas, M.\ 2004, \aap, 414, 69

\bibitem[Wilson et al.(2005)]{WILSON05} Wilson, B.~A., Dame, T.~M., Masheder,
  M.~R.~W., \& Thaddeus, P.\ 2005, \aap, 430, 523

\bibitem[Young et al.(1995)]{YOUNG95} Young, J.~S., et al.\ 1995, \apjs, 98,
219

\bibitem[Young \& Scoville(1991)]{YOUNG91} Young, J.~S., \& Scoville, N.~Z.\
  1991, \araa, 29, 581

\bibitem[Young et al.(1996)]{YOUNG96} Young, J.~S., Allen, L., Kenney,
  J.~D.~P., Lesser, A., \& Rownd, B.\ 1996, \aj, 112, 1903

\end{thebibliography}
\end{document}